\documentclass{article}
\usepackage{graphicx, cite, comment, authblk}
\title{Monetary Evolution: How Societies Shaped Money from Antiquity to Cryptocurrencies}
\author{Mahya Karbalaii}
\affil{\small{LUISS Data Lab, LUISS Guido Carli, Viale Pola, 12, 00198 Roma, Italy\\ mkarbalaii@luiss.it}}
\date{January 2025}

\begin{document}
\maketitle

\begin{abstract}
With the growing popularity and rising value of cryptocurrencies, skepticism surrounding this groundbreaking innovation persists. Many financial and business experts argue that the value created in the cryptocurrency realm resembles the generation of currency from thin air. However, a historical analysis of the fundamental concepts that have shaped money reveals striking parallels with past transformations in human society. This study extends these historical insights to the present era, demonstrating how enduring monetary concepts are once again redefining our understanding of money and reshaping its form. Additionally, we offer novel interpretations of cryptocurrency by linking the intrinsic nature of money, the communities it fosters, and the cryptographic technologies that have provided the infrastructure for this transformative shift.
\end{abstract}
{\bf Keywords:} Cryptocurrency, Blockchain Technology, History of Money, Bitcoin, Cryptography in digital currency

\section{introduction}
By tracing the history of money and its invention nearly 6000 years ago, we can observe how frequently its form has changed to remain a suitable medium for meeting the needs of societies.
Money, since its invention, has gone through many different transformations \cite{ascent-of-money,debt-5000-years, medieval1986grierson}.
This historical progression underscores that the concept of money cannot be disentangled from the communities that use it.
Indeed, the characteristics of these communities and their evolving lifestyles over millennia and centuries have consistently driven the transformation of money \cite{das2002beier, brief-history-ledgers, brief-history-of-money}.
This perspective helps illustrate that the current changes in global communities are once again reshaping the concept of money - redefining what can be considered currency and how value is assigned to it. Such transformations, however, are gradual and often span decades, as history repeatedly demonstrates.

In this study, we trace the historical evolution of the fundamental concepts that have shaped money and examine how these concepts have undergone various transformations over time. Building upon this framework, we extend the same logic to our contemporary era, illustrating how these enduring concepts are once again redefining our understanding of money and reshaping its form. Viewed through this lens, the emergence of "cryptocurrency" becomes a natural progression in the ongoing evolution of monetary systems.

Furthermore, we present novel interpretations of the concept of cryptocurrency, linking the intrinsic nature of money, the communities that form around it, and the innovative cryptographic technologies that emerged during the 1990's. These technological advances culminated in the creation of the first cryptocurrency, Bitcoin, in 2009.
Our interpretations aim to bridge the conceptual and technical aspects of cryptocurrency, offering a cohesive perspective that simplifies the understanding of its underlying mechanisms. By connecting the historical, societal, and technological dimensions of money, this study provides a comprehensive framework for comprehending the transformative impact of cryptocurrencies on modern financial systems.

The rapid increase in personal computer ownership and the expansion of Internet usage in the late 20th century became driving forces behind reforms in how money is used. This new ear led to new users' demand, such as the need for electronic payments. In parallel, one challenging question rose: how to replicate cash money in digital form. Responding to this challenge turned out to be very complicated.

Eventually, blockchain technology has emerged as a transformative force in digital systems, offering a decentralized, secure, and transparent way of recording transactions. At its core, a blockchain is a distributed ledger where cryptographically signed transactions are grouped into blocks \cite{antonopoulos2017mastering, tapscott2016blockchain, narayanan2016bitcoin, yaga2018blockchain}. 
These blocks are linked in a chronological sequence, ensuring data integrity and immutability. Each addition to the blockchain undergoes a consensus process \cite{yaga2018blockchain, bouraga2021taxonomy,abhishek2023survey}, where network participants agree on the validity of transactions, enabling trustless collaboration in environments with little or no prior trust.

This study delves into the fundamental components of blockchain systems, such as hashing, asymmetric cryptography, and distributed ledger structures, and explores their roles in achieving robust, tamper-resistant networks \cite{cachin2011introduction, yaga2018blockchain, narayanan2016bitcoin}. 
The Bitcoin ecosystem, often synonymous with blockchain, is used as a primary reference due to its foundational influence and widespread adoption. However, key concepts and mechanisms extend beyond Bitcoin to other blockchain systems, with variations aimed at optimizing efficiency, scalability, and security.

Consensus mechanisms play a pivotal role in maintaining blockchain integrity and security. Proof of Work (PoW) \cite{nakamoto2009bitcoin, abhishek2023survey, bouraga2021taxonomy}, the original consensus protocol, relies on computational puzzles that require significant energy and resources. 
While effective, its high energy consumption has driven the development of alternatives such as Proof of Stake (PoS) \cite{yaga2018blockchain, bouraga2021taxonomy, king2012ppcoin}, which selects validators based on their stake in the network rather than computational effort. These protocols ensure fairness and security while motivating stakeholders to contribute actively to maintaining the ledger.

To further elucidate the principles underpinning blockchain, this work examines practical examples of cryptographic processes like SHA-256 hashing, Merkle trees, and consensus protocols.
The Byzantine Generals Problem — a theoretical model addressing the challenges of achieving agreement in the presence of malicious actors \cite{lamport1982byzantine} — is also analyzed, providing a conceptual foundation for understanding blockchain’s resilience to attacks.

The study concludes by discussing the socio-economic incentives that encourage stakeholder participation in blockchain ecosystems, emphasizing the interplay between community trust, ledger reliability, and cryptocurrency value. 
By presenting a comprehensive overview of blockchain’s technical and theoretical aspects, this work aims to deepen the reader’s understanding of the mechanisms driving its success and adoption in various domains.

\section{Debt Recording: The Primary Driver Behind the Creation of Money}
While conventional economists believe that the bartering of goods and services led to the creation of money \cite{ascent-of-money}, anthropologists and archaeologists argue that ancient societies introduced currency to facilitate the settlement of debts \cite{debt-5000-years}.
The reconstruction of early monetary systems relies heavily on archaeological findings. Evidence suggests that over 5,000 years ago, the ancient Mesopotamians began recording quantities on clay tablets \cite{brief-history-ledgers}. The earliest known examples of Proto-cuneiform writing, dating to approximately 3200–3000 BCE, were discovered in Uruk \cite{cuneiform}.
\begin{figure}
    \centering
    \includegraphics[width=0.5\linewidth]{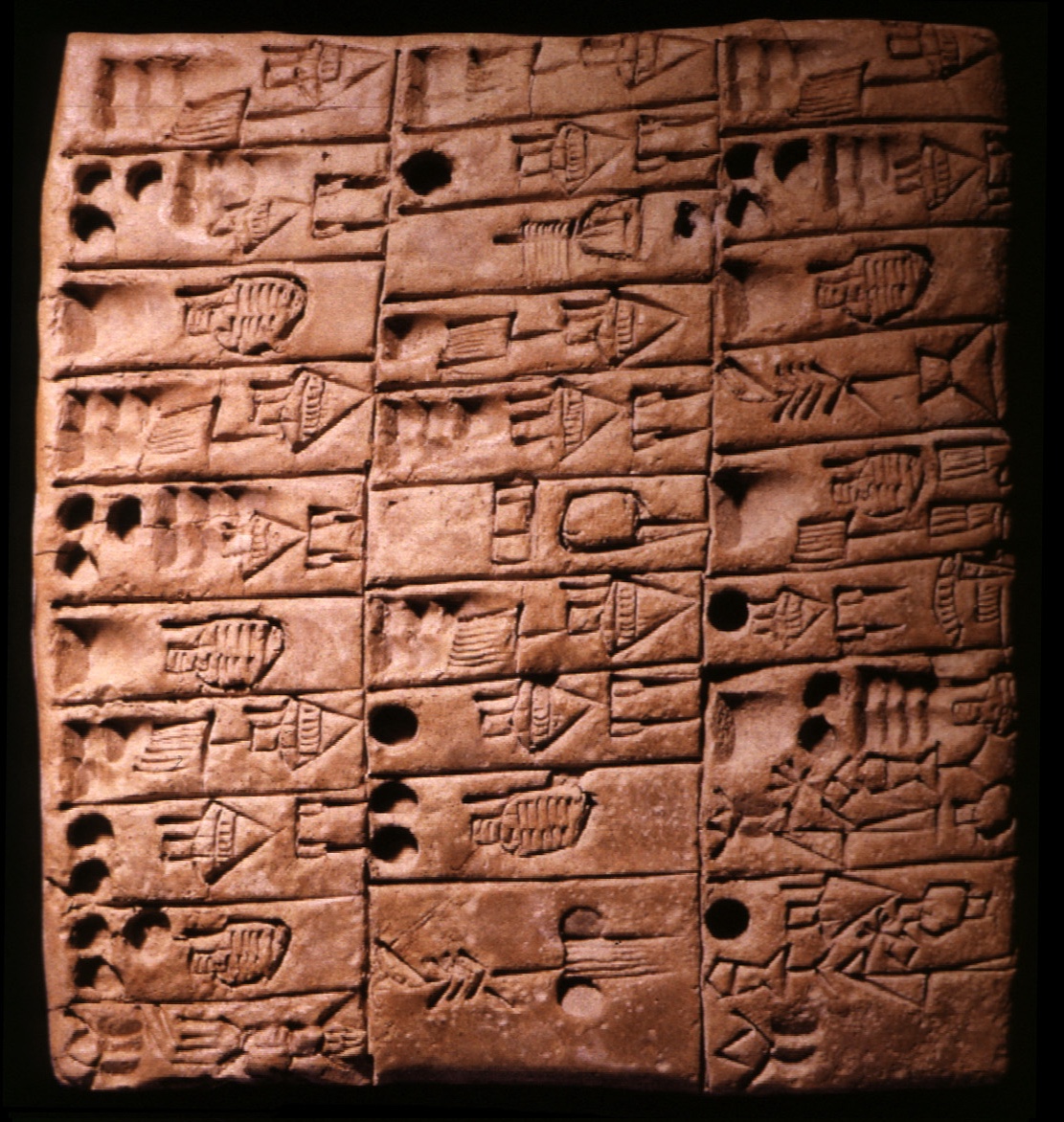}
    \caption{\textit{Administrative tablet found in Uruk, ca. 3200-3000 BC period. Now kept in Bolaffi, Turin, Italy}}
    \label{fig:fig01}
\end{figure}

These tablets (Figure \ref{fig:fig01}) primarily served administrative purposes, documenting transactions with remarkable precision. Each pictographic inscription described the nature of the transaction, including calculations, totals, and subtotals for verification. Official seals authenticated the records, which also specified the transaction date \cite{cuneiform}.

At this stage in history, money was not primarily intended as a medium for buying and selling goods. Instead, it played a crucial role in structuring social relationships and economic obligations.
Early governments utilized money to manage public debts and collect taxes. Farmers, for instance, were taxed to sustain royal households and public projects. When farmers and other commoners were unable to meet their obligations in goods, their debts were recorded in shekels.
Similarly, merchants and tradespeople often acquired goods from temple and palace officials on credit \cite{conflict-history-money}.

While economists have traditionally posited that money emerged to facilitate the bartering of goods — such as exchanging potatoes for cloth — historians argue that the advent of ledgers and record-keeping was the foundational development in this process\cite{debt-5000-years}.
The systematic recording of debts and transactions appears to have been pivotal in the evolution of monetary systems.

\section{The Evolution of Money Through\\Standardization}
The introduction of standardized metal coins marked a significant milestone in economic history. This innovation occurred in the seventh century BCE in the small kingdom of Lydia, located in present-day Turkey \cite{ascent-of-money,fountain-of-fortune}.
The Lydians produced coins from electrum, a naturally occurring gold-silver alloy. Each coin was identical in weight and composition, ensuring consistency and reliability in transactions. Unlike other forms of exchange, coins did not deteriorate with time, adding to their utility and appeal.

Following Lydia's example, major powers such as Persia, Greece, and Egypt quickly adopted hard currency. Governments recognized the advantages of standardized coins in facilitating tax collection and supporting the development of military forces.

The Roman Empire's long-standing economic success can certainly be attributed in part to the standardization of its monetary system. The denarius, which became the Roman Empire's primary currency for over four centuries, was introduced in 211 BCE. This denomination was initially minted in large quantities, with the silver required for its production likely coming from the plundering of Syracuse in 212 BCE. The denarius was valued at ten asses and was marked with the Roman numeral "X." It weighed approximately 4.5 grams, equivalent to one seventy-second of a Roman pound (libra) \cite{das2002beier}.
Two additional coin denominations were introduced alongside the denarius:
\begin{itemize}
    \item The quinarius nummus, worth half a denarius and marked with the Roman numeral "V."
    \item The sestertius, worth a quarter of a denarius and marked with the symbol "IIS."
\end{itemize}

Gold coins were not merely representations of value but were perceived as embodiments of it, owing to the intrinsic worth attributed to gold. This intrinsic value redefined society's relationship with money.
Rather than being viewed solely as a tool for social transactions, money began to be seen as an integral \textit{element of the economy}. This shift encouraged the practice of investing money to generate further wealth, laying the groundwork for more complex economic systems \cite{ascent-of-money,brief-history-of-money}.

As societies grew and interactions expanded, money also became a practical medium for transactions involving strangers or individuals unlikely to meet again, highlighting its role in enabling broader economic exchanges.

\subsection{Charlemagne's Monetary Reforms: Foundations of\\Economic Unity}

Under Emperor Charlemagne, the Frankish empire saw profound and lasting changes to its coinage system, reflecting his vision for a unified and efficient realm \cite{charlemagne2008mckitt, medieval1986grierson}. His efforts to make the empire easier to govern by standardizing external conditions extended beyond the realm of administration. They included pivotal reforms in writing, language, and education (known collectively as the Carolingian educational reform) and encompassed the standardization of systems of measurement and currency.

Charlemagne's monetary reforms were instrumental in creating a unified economic framework that transcended his reign. From 793-4 onward, he introduced the pound as the primary unit of account for coins, establishing a clear basis for monetary transactions. This reform also introduced standardized coin designs, ensuring that coins bore consistent weights and symbols, which enhanced their reliability and facilitated trade across the empire. The most commonly minted coin was the silver denarius, which became the standard currency throughout the Frankish empire.
Charlemagne's system was not just practical; it was visionary. By creating a standardized monetary system, he enabled economic cohesion in a vast and diverse territory. This effort not only improved internal commerce but also boosted trust in the currency, which became crucial for fostering trade relations with neighboring regions.

The influence of Charlemagne's monetary reforms persisted long after his reign. His system laid the foundation for medieval European coinage and influenced the development of later monetary systems across Europe. In many ways, his reforms mirrored earlier efforts at standardization, such as those of the Roman Empire with the denarius, though Charlemagne's innovations adapted these principles to the needs of a fragmented post-Roman Europe \cite{charlemagne2008mckitt, uniwuerzburg}.
Remarkably, a uniform currency area comparable to the one Charlemagne established was not achieved again until the introduction of the Euro in the late 20th century. This modern monetary union echoes Charlemagne's vision of unity through standardization, underscoring the enduring significance of his economic reforms.

\subsection{The Next Passage: Paper Money}
In the 13th century, the Chinese emperor Kublai Khan initiated a pioneering monetary reform aimed to unifying the fragmented economic landscape of his empire \cite{fountain-of-fortune}. At the time, China was divided into multiple regions, each issuing its own coinage, which hindered internal trade. To address this, Kublai Khan decreed the adoption of paper money as the principal medium of exchange.
While the concept of paper currency was not novel - earlier dynasties had introduced it in tandem with traditional coins - Kublai Khan's approach was unprecedented. Perhaps, the expanding desire in the society facilitated the acceptance of it also. The portability of paper money made it far more convenient for trade and travel, as large sums could be carried without the physical burden of metal coinage.
It was also significantly more efficient and cost-effective to produce paper money, enabling authorities to issue currency at scale without the resource constraints of precious metals.

This transformation was also captured by the Venetian merchant Marco Polo, who visited China shortly after. He expressed his astonishment by seeing a system where labor and goods were exchanged for what appeared to be mere pieces of paper. To Marco Polo, it seemed as though value was being conjured from thin air.

Kublai Khan's innovation was remarkably forward-thinking. He understood that the essence of money lies not in its physical form or intrinsic backing but in the collective trust and belief of those who use it.

\subsection{Fiat Money: A Response to Economic Fluctuations}
Although initially, paper money was backed by gold, but governments soon identified significant limitations in this system, particularly its propensity to induce deflation. As populations and economies expanded, the fixed relationship between the money supply and gold reserves created constraints. The finite nature of gold and the challenges associated with increasing its supply led to a scarcity of money, which, in turn, triggered deflationary pressures \cite{brief-history-of-money}.

To mitigate these issues, the concept of fiat money was developed, representing a paradigm shift in monetary systems. This transition moved societal trust in currency from being anchored in a tangible asset, such as gold, to reliance on the authority of the issuing entity.
Nevertheless, the underlying principle remains consistent: gold derives its value from a collective societal agreement to regard it as a medium of exchange. Similarly, fiat currency acquires its value through collective trust and acceptance, irrespective of any physical backing.

According to economists, money as the social convention and lubricant, has four main characteristics which define its main use:
\begin{enumerate}
    \item Medium to pay debts: As we saw, money was invented to pay debts.
    \item Measure for value: money allows to assign value and price to different goods and makes it possible to compare their value and they can be divided into smaller units.
    \item Medium of exchange: it permits goods and services to be exchanged with each other.
    \item Storage of value: money allows to defer consumption until a later date.
\end{enumerate}

\subsection{Internet Era and Transition To Digital Payments}
In the late 20th century, the rapid increase in personal computer ownership and the expansion of Internet usage became driving forces behind new reforms in how money is used. As more and more people began incorporating the internet into their daily lives, this new technology generated new user demands. By the 1990's, businesses recognized the growing importance of establishing an online presence to reach potential customers. However, a critical component was still missing: secure online payment channels.

Although users could browse products and services online and place orders, they often lacked the ability to pay digitally. This inconvenience hindered both users and businesses, as companies risked delivering goods without guaranteed payment. In response, the banking sector began developing solutions for secure online transactions.
The introduction of secured payment channels enabled customers to pay businesses directly using credit cards or bank accounts. However, many users remained hesitant to share personal and financial information with unfamiliar businesses, creating an opportunity for third-party payment systems. Platforms like PayPal emerged as intermediaries, providing secure and trusted transaction methods.

Over time, advancements in security systems allowed the banking sector to build sufficient trust, enabling broader adoption of online credit card transactions. This level of trust, however, was absent in the early 1990's, when the Internet was still a novelty, and users were unfamiliar with its capabilities. Establishing trust between websites and users during this period was a significant challenge.

One potential remedy to address this fear could have been the use of digital cash. However, as we will explore in the next section, replicating cash in the digital world turned out to be a highly complex technical challenge. Several initiatives from the early 1990's attempted to meet this demand and offer solutions \cite{chaum1983blind, back2002hashcash, chaum1988untraceable, haber1991time}, though with limited success. Despite the failure of many of these ventures, their efforts contributed to significant advancements in technology and knowledge.

These innovations ultimately paved the way for creation of the first digital currencies. In 2009, Bitcoin, introduced by the pseudonymous Satoshi Nakamoto, emerged as the first fully functional cryptocurrency, marking a pivotal moment in financial technology \cite{nakamoto2009bitcoin}.

\section{Digital Cash as a New Form of Money}
Although electronic payments are widely used today, it is important to note that digital payments are not synonymous with digital money, as they differ in their characteristics.
The increasing use of banking cards and online transfers has made tracing individuals' economic activities significantly easier. Online and card-based payments require users to provide identifying information, and transaction data is recorded by payment systems. This data is often analyzed to identify behavioral patterns, which third-party companies leverage to improve products and optimize marketing strategies \cite{acquisti2004privacy}.
In contrast, cash payments offer anonymity, making large-scale data collection nearly impossible. The anonymity inherent in physical cash presents an advantage that digital systems struggle to replicate. While physical cash is straightforward to produce and use, creating its digital equivalent poses substantial challenges.

\subsection{The Problem of Double Spending}
Attempts to replicate physical cash in digital environments and use it on digital platforms introduced a critical challenge: "double spending." Since translating any real-world asset into the digital realm requires creating a "data" representation of it, replicating cash was no exception. However, making multiple copies of this "piece of data" is far easier than counterfeiting a physical banknote. This issue, known as the "double-spending problem," became one of the major obstacles to introducing digital cash, as a single digital token could be duplicated and spent multiple times, thereby undermining its value \cite{chaum1983blind}.

Physical cash relies on authenticity rather than the payer’s identity; the recipient only needs assurance that the cash is genuine. Digital cash, however, is represented as data packets, which can be easily copied. Without robust mechanisms, malicious actors could reuse digital tokens, leading to systemic issues.
The first notable solution to the double-spending problem was introduced by David Chaum in 1983. Chaum’s pioneering work applied cryptographic techniques to ensure the authenticity and value of digital cash \cite{chaum1983blind}. His system involved issuing digital cash with unique serial numbers, verified by a trusted entity. This issuer, not necessarily a bank, served as an intermediary to authenticate transactions.
The transaction process in Chaum’s system ensured security as follows: 
\begin{enumerate}
    \item Each unit of digital cash carried a unique serial number issued by the trusted entity.
    \item During a transaction, the recipient could verify the serial number with the issuer to confirm it had not been spent.
    \item If valid, the transaction proceeded, and the serial number was marked as spent in the issuer’s ledger. A new serial number was generated for the recipient, ensuring the note’s continued validity and preventing reuse.
\end{enumerate}
This approach addressed the double-spending problem but had limitations:
\begin{itemize}
    \item \textbf{Loss of anonymity}: Unlike physical cash, verifying transactions required identifying and authenticating users, compromising privacy.
    \item \textbf{Centralized trust}: The system depended heavily on a central authority, introducing a single point of failure.
\end{itemize}

The system’s reliance on real-time ledger access and secure data management made it cumbersome and inefficient. Despite these challenges, Chaum’s work laid the foundation for subsequent advancements in electronic cash systems \cite{chaum1988untraceable}.
These innovations eventually led to decentralized solutions, such as Bitcoin, which eliminated the need for centralized trust while addressing the double-spending problem through blockchain technology.
As Chaum and others struggled to address the technical challenges of the double-spending problem in an efficient way, the development of new cryptographic tools — initially designed to address entirely different issues — ultimately gave rise to groundbreaking technologies.

\subsection{Digital Money for Digital Effort} \label{Digital Money for Digital Effort}
As techniques to replicate physical cash in the digital realm advanced, a pivotal question emerged: could digital efforts be rewarded with digital money? If the logic underlying the issuance of fiat money could be reproduced digitally, would it be possible to create an entirely new form of currency in cyberspace? This idea inspired early proposals like NetCash and e-Gold in the 1990's \cite{chaum1983blind, goldschlag1996netcash}. While these initiatives failed to materialize into practical systems, they laid the groundwork for further exploration.
The concept of assigning value to digital objects dates back to 1992, when cryptographers Cynthia Dwork and Moni Naor proposed that solving computational puzzles could hold economic value \cite{dwork1992pricing}. They introduced this idea to combat spam emails by requiring senders to solve cryptographic puzzles as proof of effort. This concept later evolved and became known as "proof-of-work." Simply put, solving such computationally intensive problems could only be done if the user had invested enough effort—i.e., work—to find the solution. In other words, providing the "solution" to the puzzle would serve as proof of the effort expended.

Around the same time, two other cryptographers, Ron Rivest and Adi Shamir, known for their contributions to the RSA cryptosystem, introduced computationally intensive puzzles as a form of proof of work \cite{rivest1996micropayments}. Despite differing implementation methods, these systems shared a core principle: computational puzzles whose solutions could hold value. These puzzles had to meet specific conditions:
\begin{enumerate}
    \item \textbf{Dependency on inputs}: Each puzzle must rely on unique inputs such as content, time, or transaction data, ensuring solutions are non-reusable.
    \item \textbf{Easy verification}: The solution must be easily verifiable by recipients without repeating the computational process.
    \item \textbf{Independence}: Solving one puzzle should not make solving others any easier.
\end{enumerate}

Building on this, Adam Back developed Hashcash in 1997, a system that assigned value to computational work \cite{back2002hashcash}.
Adam Back’s Hashcash successfully demonstrated these principles in the context of spam email detection, but it never achieved significant adoption. One likely reason was the rapid development of alternative solutions. As new methods emerged, email service providers found it easier to filter spam directly from users’ inboxes without requiring computational puzzles. As a result, there was little incentive to reward \emph{digital effort} with \emph{digital cash} \cite{goodman2003spam}.

Ultimately, as with previous ideas, the lack of a sufficiently large and committed community prevented these systems from evolving into successful forms of digital money. Despite their limited success, these early systems provided critical insights into the potential of computational effort as a basis for digital currency.

\section{Pivoting the Problem: From User's Identity to Currency's Identity}
As previously discussed, the two primary characteristics of physical cash are the preservation of anonymity, particularly the identity of the spender, and the assurance of authenticity, meaning that the note or coin represents a legitimate and recognized value. In the context of electronic payments, transactions are typically validated when the payee receives confirmation from the payer's bank that sufficient funds are available to complete the transaction. However, this process inherently requires revealing the payer's identity to at least some level of authority, making it incompatible with the desire for full anonymity.

A potential solution to this challenge lies in shifting the focus from verifying the payer's identity to verifying the "digital money identity." Instead of validating the payer's identity and account details, the payee could authenticate the digital money itself to determine whether it has been previously used. This approach involves verifying the unique identification of the digital money and ensuring it has not already been spent, effectively addressing concerns about double-spending.

This verification process can be implemented by creating an immutable, open-source ledger accessible to all participants in the community. Such a ledger would record the unique identification numbers or codes of all previously used digital coins. When a payee receives a digital coin, they can efficiently verify its history by checking the ledger. If the coin has already been used, it would no longer carry the indicated value, and the payee could reject the transaction. Conversely, if the coin is verified as unused, the payee can accept it for payment. The transaction would then be recorded in the ledger, assigning a new identification code to the coin for the new holder. Importantly, this process is timestamped, ensuring the chronological integrity of all transactions.

This concept forms the foundation of the solution proposed by Satoshi Nakamoto in his seminal work \cite{nakamoto2009bitcoin}. However, implementing this solution required the development of advanced technologies, particularly cryptographic techniques. By integrating these techniques, many of which were pioneered by independent initiatives, Nakamoto successfully addressed the double-spending problem while preserving the anonymity of the payer.

\subsection{Distributed Ledger Based on Blockchain Technology}
Bitcoin’s creator, Satoshi Nakamoto, synthesized all essential components into a cohesive system by introducing the blockchain — a distributed ledger accessible to any member of the community for recording all past transactions. Nakamoto’s proposal emphasized that this distributed ledger should possess specific characteristics to ensure its reliability, security, and transparency.
\begin{enumerate}
    \item \textbf{User anonymity}: Identities are encoded and disclosed only at the users' discretion. This builds on Chaum’s work on selective disclosure and public-private key encryption \cite{chaum1988untraceable}.
    \item \textbf{Immutability}: The ledger is immutable; no block can be removed, nor can any information within a block be altered. Hashing functions ensure that even minor changes to the ledger's content produce completely different outputs, making tampering immediately detectable \cite{damgard1989design}. A key property of hashing functions is the avalanche effect: even a minor change to the input—no matter how small—produces a completely different output. This ensures that two distinct inputs, even if highly similar, will yield entirely different outputs. As a result, any attempt to tamper with the data becomes immediately apparent, reducing the risk that such changes could go unnoticed.
    \item \textbf{Append-only structure}: Blockchain technology, initially introduced by Haber and Stornetta in 1991, ensures new data can only be added sequentially, preserving the chronological order of transactions \cite{haber1991time}. The timestamps accurately reflect the order of creation, preserving the sequence of events. If one document is created before another, the timestamps will show this order definitively.
    \item \textbf{Distributed trust}: The ledger is replicated across the network, eliminating reliance on a central authority. By distributing the ledger across a network of participants and allowing open access to all transactions, the need for a central trusted authority is eliminated. Trust is distributed among participants who collectively validate transactions \cite{narayanan2016bitcoin}.
\end{enumerate}
Blockchain technology thus became the cornerstone of Bitcoin’s success, providing a robust framework for decentralized and anonymous electronic payments.
In other words, much like 6,000 years ago when ledgers served as the foundational building blocks for the creation of money, this innovation has succeeded in generating a new form of currency, known cryptocurrency.

\section{The Core Components of Blockchain\\Technology}
Blockchains are distributed, decentralized digital ledgers designed to securely record transactions. Each transaction is cryptographically signed and grouped into blocks. Once validated and agreed upon through a consensus mechanism, each block is linked cryptographically to its predecessor, forming a chain. 
As new blocks are added, the older blocks become increasingly difficult to alter, ensuring immutability. The blockchain ledger is replicated across all participating nodes in the network, with conflicts resolved automatically using predefined rules. This section explores the fundamental elements of blockchain technology, with a focus on the Bitcoin ecosystem \cite{antonopoulos2017mastering, tapscott2016blockchain, narayanan2016bitcoin, yaga2018blockchain}.
While many features discussed are common across blockchain platforms, variations exist to enhance specific attributes like efficiency, often at the cost of others.

\textbf{Users and Nodes in Blockchain Systems}

Users in the blockchain systems are entities (individuals, organizations, businesses, or governments) utilizing the blockchain for transactions or data management.
Nodes are the building blocks of the blockchain network and come in three primary forms:
\begin{itemize}
    \item Full Nodes: Store the entire blockchain and validate transactions independently.
    \item Mining Nodes: Specialized full nodes responsible for solving cryptographic puzzles and publishing new blocks.
    \item Lightweight Nodes: Operate without storing the full blockchain, relying on other nodes for data verification.
\end{itemize}

\subsection{Cryptographic Techniques in Blockchain}
Hash functions play a crucial role in blockchain operations. These are cryptographic algorithms that generate a fixed-size output (hash) from an input of arbitrary size.
Key properties of hash functions include:

\textbf{One-Way Functionality:} It is computationally infeasible to reverse\\-engineer the input from the hash.

\textbf{Collision Resistance:} It is nearly impossible to find two distinct inputs that produce the same hash.

Bitcoin employs a combination of two hashing algorithms (notably SHA-256) to enhance security against attacks. 
In contrast, permissioned blockchains often use less computationally intensive hashes, given their reduced need for robust consensus mechanisms.

\subsection{Asymmetric-Key Cryptography}
Blockchain systems leverage asymmetric-key cryptography for secure transactions. 
Each user generates:

\textbf{Private Keys:} Kept secret and used to sign transactions, proving ownership of digital assets.

\textbf{Public Keys:} Shared openly and used to verify signatures or derive block\\-chain addresses.

For security, it is recommended to use each key pair only once. Transactions assigning assets utilize the public key, while subsequent transactions must be verified with the corresponding private key.

\subsection{Blockchain Ledger Types}
Unlike centralized ledgers maintained by a single authority, blockchain ledgers are decentralized and distributed across the network. 

Reliability is achieved through consensus mechanisms \cite{cachin2011introduction, decker2013information}. Bitcoin employs a transaction ledger, recording all individual transactions, whereas other systems, like Ethereum, use balance ledgers.

\subsection{Structure of a Blockchain Block}
Each block in a blockchain contains the following elements:
\begin{enumerate}
    \item Block Number (Height): The position of the block in the chain.
    \item Previous Block Hash: Cryptographic link to the preceding block.
    \item Current Block Hash: Hash of the block's content.
    \item Merkle Tree Root: Summary hash of all transactions in the block.
    \item Timestamp: Time the block was created.
    \item Block Size: Data size of the block.
    \item Nonce: A number adjusted by miners to solve the cryptographic puzzle.
    \item Transaction List: Details of transactions included in the block.
\end{enumerate}

\subsection{The Nonce} \label{pow-example}
The nonce, often thought to derive from "number used once," is a key element in mining. Miners adjust the nonce to find a hash value that satisfies specific criteria, such as starting with a predetermined number of zeros. 
This process, part of the Proof-of-Work consensus, ensures secure and decentralized block creation \cite{nakamoto2009bitcoin, boehme2015bitcoin}.

Example of PoW Puzzle Solving
Consider a mining puzzle where miners must find a hash value starting with "000000" using the SHA-256 algorithm: SHA256("blockchain" + Nonce) = Hash Value starting with six leading zeros.

\[\ Input: Blockchain0\ (not solved)\]
\[\ Nonce: 0\]
\[\ Hash: 0xbd4824d8ee63fc82392a6441444166d22ed84eaa6dab11d4923075\]
\[\ 975acab938\]
\bigbreak

\[\ Input: Blockchain1\ (not solved)\]
\[\ Nonce: 1\]
\[\ Hash: 0xdb0b9c1cb5e9c680dfff7482f1a8efad0e786f41b6b89a758fb26d\]
\[\ 9e223e0a10\]
\bigbreak

\[\ Input: Blockchain10730895\ (solved)\]
\[\ Nonce: 10730895\]
\[\ Hash: 000000ca1415e0bec568f6f605fcc83d18cac7a4e6c219a957c10c\]
\[\ 6879d67587\]
\[\ Time taken: 5.04 seconds\]

In this case, it takes 10,730,895 guesses in about 5.04 seconds to find the correct nonce using Macbook Air M1 with 8 GB Ram.
Each additional leading zero increases the difficulty exponentially. For instance, with one additional zero, i.e. 7 leading zeros, the same hardware required 460.82 seconds to make 934 million guesses to solve the puzzle.

Benchmark for 7 leading zeros:
\[\ Input: Blockchain10730895\ (solved)\]
\[\ Nonce: 934224174\]
\[\ Hash: 0000000e2ae7e4240df80692b7e586ea7a977eacbd031819d0e603\]
\[\ 257edb3a81\]
\[\ Time taken: 460.82 seconds\]

And for one more leading zero, the same machine required 4183.97 seconds, i.e. nearly 830 times longer to guess 8795718656 numbers and solve the puzzle.

Benchmark for 8 leading zeros:
\[\ Input: Blockchain8795718656\ (solved)\]
\[\ Nonce: 8795718656\]
\[\ Hash: 0000000041095df5b11e4775bac1a087d3eaeffc15ff0bb7b5c3dd\]
\[\ aecb4beb64\]
\[\ Time taken: 4183.97 seconds\]

Despite the effort required, verifying a nonce is computationally simple, as only one hash calculation is needed to confirm its validity.

\section{Mining Blocks, Earning Rewards}
Bitcoin’s success lies in incentivizing participants to maintain and update the ledger securely. This process, known as mining, involves solving complex computational puzzles tied to the creation of new blocks in the blockchain. These puzzles exhibit characteristics like dependence on block content and time, ensuring their uniqueness.

Each new block in the Bitcoin blockchain contains the details of all recent transactions, additional data linking it to the preceding block, and further information. To ensure the integrity of this information, specific nodes in the network—referred to as miners—are tasked with solving a complex computational puzzle. Solving this puzzle requires substantial computational resources. The computational problem is designed to possess specific characteristics, such as being tightly dependent on the content of the block, the time of generation, and other factors.

Once a miner solves the puzzle, the solution is easily verified by the network. Upon successful validation, miners are rewarded for their efforts with newly minted Bitcoin. As of December 2024, the reward for mining a new block is 3.5 Bitcoins \cite{bitcoinreward2024}. This mechanism aligns incentives, ensuring computational effort secures the network and sustains the ecosystem \cite{bonneau2015bitcoin}.

\subsection{Bitcoin's PoW and Fluctuating Puzzle Difficulty}
Bitcoin’s PoW system is designed for trustless environments where participants do not inherently trust each other \cite{garay2015backbone, narayanan2016bitcoin}. However, the process is computationally expensive, with significant environmental implications.

As of December 2024, Bitcoin’s hash rate reached approximately 800 exa-hashes per second (EH/s) \cite{coin2024bitcoin}, meaning miners collectively performed 800 quintillion guesses per second.
The energy consumption of the Bitcoin network was estimated to be comparable to half of Italy’s annual electricity usage.
The difficulty of mining puzzles adjusts dynamically based on the number of participating nodes.

For example: In 2023, Bitcoin required miners to generate hash values with 19 leading zeros to mine a block.
This adaptability ensures consistent block creation times, maintaining network stability as participation levels fluctuate.
By combining cryptographic techniques, distributed systems, and consensus mechanisms like PoW, blockchains provide a secure and transparent framework for digital transactions and data management. Further advancements continue to address efficiency and scalability challenges while preserving decentralization.

\section{The Role of Consensus Mechanisms}
One of the most significant features of blockchain-based applications is their ability to facilitate business transactions between untrusted and unknown parties. Even in such decentralized environments, where complete anonymity may not always be present and most participants remain unfamiliar with one another, these systems can establish a network that provides a trustworthy, tamper-resistant ledger shared among multiple nodes. The key mechanism enabling this functionality is known as "consensus."

Consensus mechanisms ensure agreement among distributed nodes about the state of the blockchain. While various blockchain platforms employ different consensus algorithms to incentivize participation, their overarching goal remains the same: to provide a distributed ledger that operates fairly, securely, and efficiently. Immutability—the guarantee that past records cannot be altered—is a foundational principle underpinning blockchain technology.
This section explores the role of consensus in cryptocurrencies, with a comparative analysis of blockchain systems and traditional fiat money.

\subsection{The Role of Central Authority in Fiat Money Systems}
In traditional financial systems, central authorities such as central banks manage the supply and circulation of fiat money. These institutions maintain trust in the currency by implementing anti-counterfeiting measures and ensuring its security within the economy. While counterfeiting remains a potential threat, central banks continuously improve physical and digital safeguards to minimize the risk of fake currency entering circulation at scale.

The legitimacy of any currency, whether fiat or digital, is inherently tied to the trust and acceptance of a community. A currency gains its value and function as a medium of exchange only when it is recognized and trusted by its users. For cryptocurrencies, this trust hinges on the security and reliability of the underlying system.

\subsection{Creating Trust in Decentralized Ledgers}
Cryptocurrencies achieve trust through decentralized and distributed ledgers. These ledgers record all transactions transparently, allowing all participants in the community to view and verify them. The underlying technology employs cryptographic methods, particularly hashing and blockchain structures, to securely link each block of data to its predecessor. This ensures that transactions are registered, validated, and stored immutably in the ledger.

The task of creating, maintaining, and validating such a ledger is complex and resource-intensive. Unlike centralized systems with a single authority, decentralized blockchain networks require participants to collectively take responsibility for the ledger’s upkeep. Consequently, a significant challenge is motivating these participants to contribute to the network’s security and functionality.

\subsection{Incentivizing Members to Increase Participation}
To replace the role of a central authority, blockchain systems rely on consensus mechanisms. These algorithms motivate participants to validate transactions, secure the ledger, and ensure the system operates efficiently. Participation in maintaining the ledger is often rewarded, creating an economic incentive to contribute \cite{nakamoto2009bitcoin, buterin2014next, kwon2014tendermint, micali2017algorand}.

Key objectives of consensus mechanisms include:
\begin{enumerate}
    \item \textbf{Security:} Ensuring that all recorded transactions are valid and protected against fraudulent activities.
    \item \textbf{Efficiency:} Validating transactions quickly to maintain user interest and system reliability.
    \item \textbf{Scalability:} Accommodating as many transactions as possible without compromising system performance.
    \item \textbf{Equilibrium:} Preventing imbalances among participant groups and mitigating vulnerabilities to attacks.
\end{enumerate}

Despite these goals, achieving an ideal balance among security, speed, and decentralization often involves trade-offs. Different blockchain platforms adopt varied consensus mechanisms to address these challenges.

\subsection{Types of Ledgers in Blockchain Systems}
There are three primary types of ledgers in the world of cryptocurrencies: public, private, and consortium ledgers \cite{yaga2018blockchain, narayanan2016bitcoin, bouraga2021taxonomy, abhishek2023survey}.
\begin{itemize}
    \item {Public Ledgers:} In public ledgers, transactions are transparent and accessible to all participants, but the identities of participants remain anonymous. These systems prioritize openness and decentralization.
    \item {Private Ledgers:} Private ledgers involve a closed group of designated participants who are authorized to validate and record transactions. While the identities of participants are known, transactions remain encrypted and private, making these ledgers suitable for organizations requiring confidentiality.
    \item{Consortium Ledgers:} Consortium ledgers are partially private systems operated under the governance of a group rather than a single entity. These ledgers combine aspects of public and private systems, balancing transparency with control.
\end{itemize}
These ledger types broadly correspond to the two main categories of blockchains: permissionless and permissioned.
Permissionless Blockchains are the ones in which any node or participant can read and register transactions, fostering openness and decentralization \cite{pass2017hybrid}.
On contrast, in permissioned Blockchains Only certain members are authorized to record transactions, enhancing efficiency and control but introducing an element of centralization.

\subsection{Balancing Decentralization and Efficiency}
Critics argue that the selective privileges granted in permissioned blockchains can undermine the original goal of avoiding central authority. However, in practice, achieving efficiency while maintaining security often necessitates such trade-offs \cite{kiayias2016speed, decker2013information, cachin2011introduction}.

Permissioned blockchains typically require less computational power, as the cryptographic puzzles involved are less intensive. While this reduces resource consumption, it can also create vulnerabilities. For instance, malicious actors within a permissioned system may exploit its reduced complexity unless additional safeguards are implemented.

Consequently, permissioned blockchains are often more suited for business applications requiring controlled environments rather than as widely accepted currencies. Ensuring robustness against attacks in such systems requires implementing strict governance and security measures.

\section{Consensus Mechanisms in Practice}
The first and most well-known consensus mechanism is Proof of Work (PoW), introduced by Bitcoin.

PoW involves solving complex mathematical puzzles to validate transactions, which requires significant computational power and energy consumption. In section \ref{pow-example}, we explained how this mechanism works.

One of the important characteristics of this consensus is that solving one puzzle does not make it easier to solve the next one. This is fundamentally important to make solving each puzzle independent of another one.
While highly secure, this mechanism has been criticized for its environmental impact.

In response to PoW’s limitations, alternative mechanisms have emerged. For example, Ethereum, initially based on PoW, transitioned to Proof of Stake (PoS) to improve energy efficiency. PoS prioritizes participants who hold more coins, granting them greater responsibility in the validation process.

\subsection{Proof of Stake (PoS): A Fair and Sustainable Consensus Mechanism}
Proof of Stake (PoS) is designed primarily for permissioned public distributed ledgers and operates on economically bonded puzzle solutions \cite{bentov2014proof, kiayias2017ouroboros}. 
Unlike Proof of Work (PoW), PoS does not involve generating new coins. Instead, miners (or validators) are rewarded solely through transaction fees, eliminating the need for block rewards.

The PoS protocol is inherently fair, as the probability \textit{p} of a participant being selected as a validator is directly proportional to the fraction \textit{p} of the total stake they own in the system. 
The process of selecting a validator for block creation is randomized, resembling a lottery system.

A question often arises: in this framework, why would wealthy stakeholders be motivated to participate in the ecosystem and actively contribute to block validation?
The answer lies in the intrinsic relationship between the currency’s value and the trust of its community. If stakeholders lose interest in validating transactions, the reliability of the ledger may degrade. 

This decline in trust could lead to diminished adoption and usage, causing the cryptocurrency’s value to drop. Since the wealth of large stakeholders is tied to the currency’s value, they have a vested interest in maintaining the ledger’s integrity. By actively validating transactions, they ensure the system remains secure, the currency retains its value, and they continue earning transaction fees.

To enhance participation, various consensus mechanisms incorporate unique strategies to incentivize community engagement \cite{pass2017hybrid, abhishek2023survey, kwon2014tendermint}. Increased transaction activity not only improves the reliability of the ledger but also enhances the visibility and popularity of the corresponding cryptocurrency, fostering its growth and adoption. Other notable consensus mechanisms include:

\textbf{Delegated Proof of Stake (DPoS):} Participants elect a smaller group of representatives to validate transactions on their behalf, enhancing efficiency while maintaining decentralization.

\textbf{Proof of Authority (PoA):} A reputation-based system where trusted validators are pre-selected, offering faster transaction processing.

\textbf{Byzantine Fault Tolerance (BFT) Variants:} Mechanisms like Practical BFT (PBFT) ensure agreement among nodes even in the presence of malicious actors. See Section \ref{byzantine}

\section{Byzantine Fault Tolerance and the Role of Consensus in Mitigating Malicious Behavior} \label{byzantine}
One of the critical challenges in blockchain networks is addressing malicious behavior within the system. A robust consensus mechanism must effectively counteract this threat to maintain the network’s security and functionality \cite{lamport1982byzantine, lamport1998part}.

A pivotal algorithm in addressing this issue is derived from the Byzantine Generals Problem \cite{lamport1982byzantine}, a concept introduced by Leslie Lamport, Robert Shostak, and Marshall Pease. This problem conceptualizes a scenario in which distributed nodes in a network must reliably communicate despite the presence of malicious participants.

\subsection{The Byzantine Generals Problem}
The problem is framed as follows: "An army of Byzantine generals, each commanding a portion of their forces, is encamped around an enemy city \cite{lamport1982byzantine}. The generals can communicate only through messengers and must agree on a common battle plan. However, one or more generals may be traitors, attempting to sow confusion. The challenge is to design an algorithm that ensures all loyal generals reach a consensus despite the presence of traitors."

To address this, the authors propose a reliable algorithm based on the principles of secure communication through writing and signature mechanisms. The algorithm can function in a system with 3m nodes, where at most m nodes may act maliciously. Its assumptions are:

Signature Integrity: The signatures of loyal participants cannot be forged, and any alteration to the content of a signed message can be detected. This is achieved using hashing mechanisms.

Authenticity Verification: The authenticity of a participant’s signature can be independently verified, a feature mirrored in blockchain’s cryptographic verification process.

In practice, each message consists of its content (e.g., an "order") and a sequence of signatures from all prior recipients. While traitors can attempt to alter the content, they cannot forge or modify the signatures of loyal participants, as such actions would be detectable.

\subsection{Practical Scenarios}
To better understand the algorithm, consider the following scenarios:

\textbf{Scenario A:} Loyal Commander, One Traitor Lieutenant

\begin{figure}
    \centering
    \includegraphics[width=0.5\linewidth]{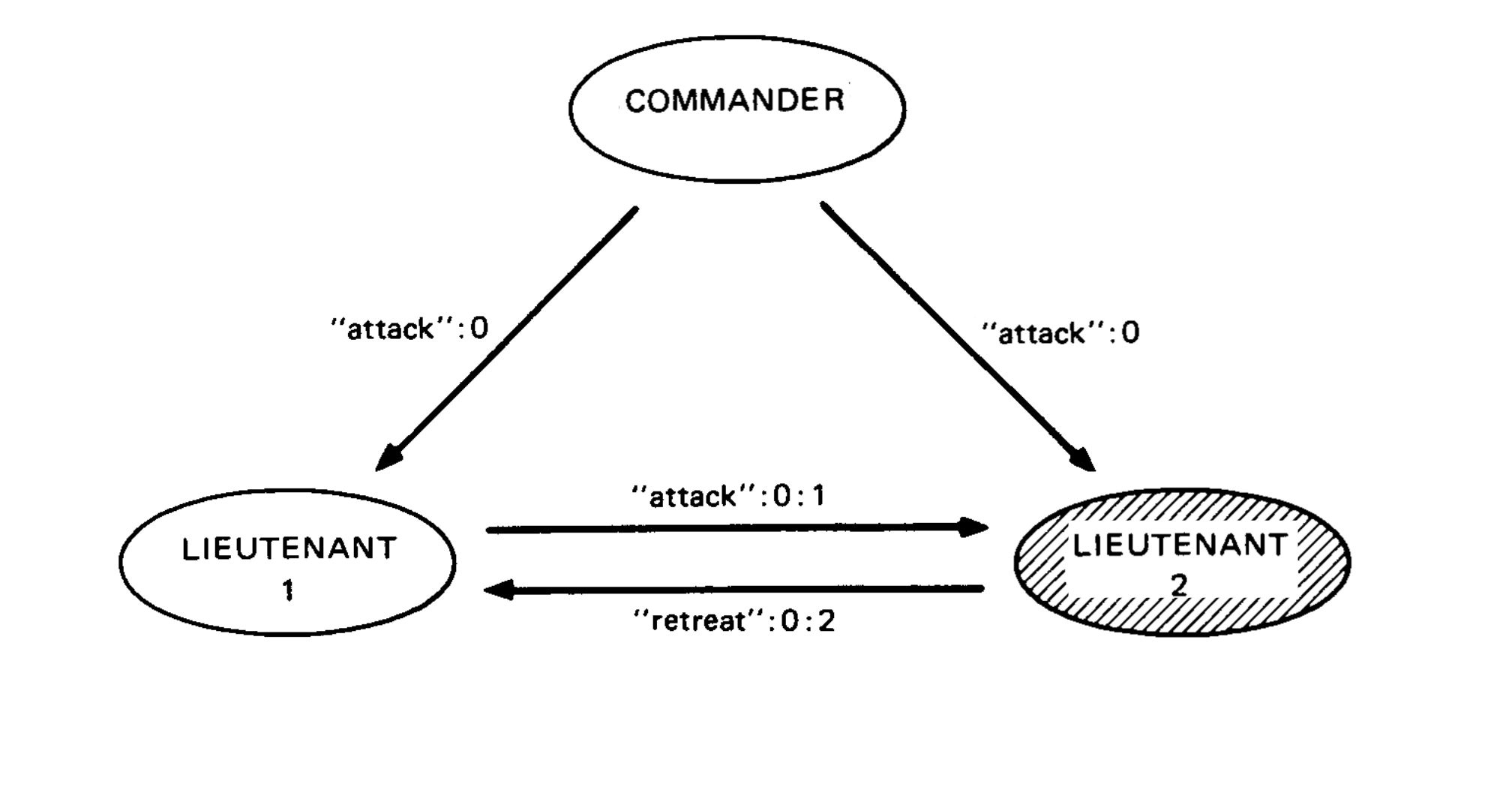}
    \caption{\textit{Loyal Commander, One Traitor Lieutenant}}
    \label{fig:loyal-commander}
\end{figure}

Three generals are involved: a loyal commander, a loyal lieutenant (Lieutenant 1), and a traitorous lieutenant - Lieutenant 2 (Figure \ref{fig:loyal-commander}).

The commander sends an identical order to both lieutenants. However, Lieutenant 2 alters the order and forwards the modified message to Lieutenant 1, along with the original signature chain.
Lieutenant 1, upon receiving both messages, observes a discrepancy. Since the commander’s signature cannot be falsified, Lieutenant 1 can identify the modified message as invalid, recognizing the traitorous behavior of Lieutenant 2.

\textbf{Scenario B:} Traitorous Commander, Loyal Lieutenants

\begin{figure}
    \centering
    \includegraphics[width=0.5\linewidth]{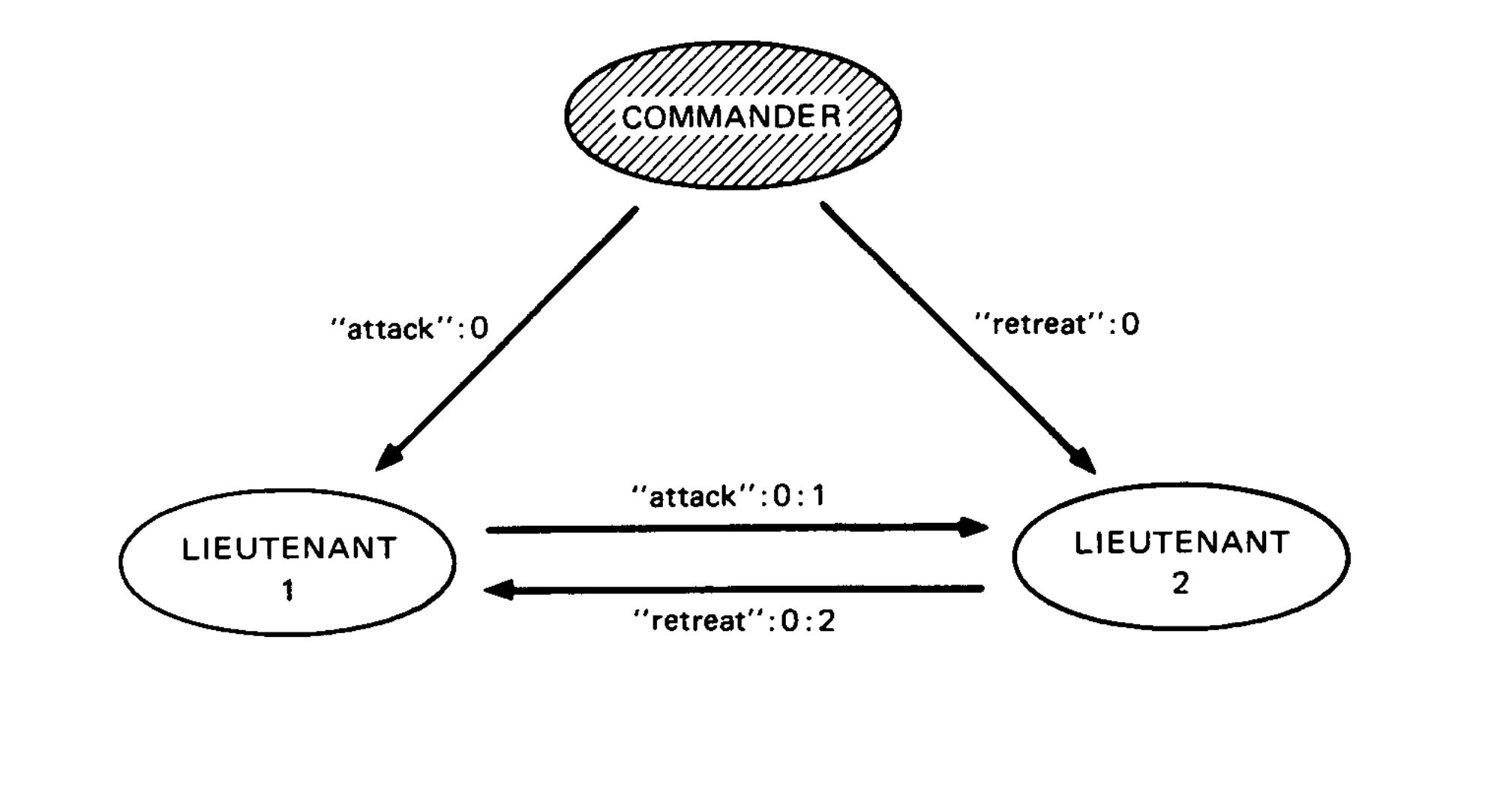}
    \caption{\textit{Traitorous Commander, Loyal Lieutenants}}
    \label{fig:traitorous-commander}
\end{figure}

Here, the commander is the traitor, while both lieutenants are loyal (Figure \ref{fig:traitorous-commander}).
The commander sends conflicting orders to the two lieutenants: one to attack and the other to retreat. Each lieutenant forwards the received order, appending their signature to the message.

When the lieutenants exchange messages, each detects a conflict between the commander’s original order and the forwarded message. Since the signature of a loyal lieutenant cannot be altered, both deduce that the commander has deliberately issued conflicting orders. Thus, they identify the commander as the traitor.

\subsection{Generalizing the Algorithm}
The authors extend this algorithm to all written communications in networks where the number of malicious participants is less than one-third of the total nodes. By leveraging secure signature mechanisms and ensuring message integrity, the system can tolerate malicious behavior up to this threshold, maintaining network reliability.

This approach forms the foundation of Byzantine Fault Tolerance (BFT), a critical concept in blockchain consensus mechanisms \cite{micali2017algorand, lamport1998part, eyal2014majority}. BFT ensures that even in the presence of malicious actors, the network can achieve consensus and remain secure.

\section{Initial Coin Offerings to Acquire Crypto\\-currencies}
As cryptocurrencies gain popularity, various methods have emerged for acquiring them. One widely-used method for obtaining cryptocurrencies is through Initial Coin Offerings (ICOs).
An Initial Coin Offering (ICO) is a fundraising method used by companies, typically in the blockchain or cryptocurrency space, to raise capital by offering tokens or coins in exchange for investments. These tokens are often seen as digital assets and can represent a range of different things, including a share in the company, access to a product or service, or participation in a decentralized network. ICOs are similar to Initial Public Offerings (IPOs) in the stock market, but instead of offering shares of a company, ICOs sell digital tokens.

While ICOs can provide opportunities for early-stage investments in innovative projects, they are also speculative, and the risks of fraud or failure are high. ICOs are not always regulated in the same way as traditional securities offerings, which adds an additional layer of risk for investors. Depending on the jurisdiction, some ICOs may be subject to regulation, while others might not.

As with many emerging trends, the increasing public interest in ICOs has unfortunately attracted fraudsters looking to exploit less informed participants.
Before deciding to participate in an ICO, it is crucial for potential investors to fully understand both the benefits and risks associated with the specific project or investment, particularly by examining the technical details provided in the-so-called "white-papers". Investors should take the time to ask detailed questions of the issuer and verify all information from independent, reliable sources. Additionally, it is important for investors to ensure that the project’s characteristics align with their personal investment goals and risk tolerance.

Regulatory bodies such as BaFin have published expert articles \cite{crypto2024bafin} on ICOs and their associated risks, offering critical insights for investors. More information on blockchain technology and virtual currencies can also be found under the section on Company Start-ups and Fintech Companies. Furthermore, national supervisory authorities and the European Securities and Markets Authority (ESMA) provide additional resources \cite{finalrep2024esma} and warnings to help investors make well-informed decisions in this rapidly evolving space.

\section{Conclusion}
The history of money reveals a continuous evolution, shaped by the changing needs and characteristics of societies throughout time.
From ancient bartering systems to the development of coinage, paper money, and now digital currencies, the concept of money has consistently adapted to facilitate trade and reflect societal trust. 

As we witness the rise of cryptocurrencies today, it is evident that money’s transformation is an ongoing process, driven by technological advancements and shifts in collective belief.
The trajectory suggests that, as with previous innovations, the future of money will be defined not by its physical form, but by the widespread acceptance and trust of the communities that adopt it.
Just as Marco Polo was astonished to witness the use of paper money in 13th-century China, many contemporary business leaders remain skeptical about the emerging notion of cryptocurrencies.

Yet, as more individuals join the growing communities that collectively agree to assign value to cryptocurrencies, their role as a medium of exchange strengthens.
Over time, it is likely that these communities will reach a scale capable of imposing their preferred form of currency, further driving the evolution of money in response to societal needs.

The success of a cryptocurrency depends on the strength and trust of its community. Without widespread acceptance, a currency is unlikely to achieve meaningful circulation. Conversely, a robust and supportive community enhances the currency’s stability, longevity, and utility as a medium of exchange. Consensus mechanisms play a pivotal role in fostering this trust. By ensuring the security and reliability of the blockchain, they provide the foundation for a vibrant, sustainable cryptocurrency ecosystem.

\section{Acknowledgments}
The author would like to express her sincere gratitude to Dr. Johannes Engels and Dr. Hamid Reza Sepangi for their careful review of the content of this article. Their valuable feedback and insightful suggestions greatly enhanced the quality of this work.

\bibliography{review-biblio}
\end{document}